\documentclass[showpacs,twocolumn,preprintnumbers,amsmath,amssymb,latexsym,array,enumerate,letter]{revtex4-1}
\usepackage{graphicx}
\usepackage{dcolumn}
\usepackage{bm}
\usepackage{amssymb}
\usepackage{epsfig,color}
\usepackage{slashed,hyperref}
\newcommand{\gbs}{g_q}
\newcommand{\gV}{g_\mu}

\newcommand{\half}{\frac{1}{2}}
\newcommand{\be}{\begin{equation}}
\newcommand{\ee}{\end{equation}}
\newcommand{\beq}{\begin{eqnarray}}
\newcommand{\eeq}{\end{eqnarray}}

\definecolor{verde}{cmyk}{0.92,0,0.59,0.25}
\definecolor{rossos}{cmyk}{0,1,1,0.55}
\definecolor{blus}{cmyk}{1,1,0,0.6}
\def\arXiv#1{{\color{rossos}\href{http://arxiv.org/abs/#1}{arXiv:#1}}} 
\hypersetup{colorlinks,bookmarksopen,bookmarksnumbered,
linkcolor=blus,pdfstartview=FitH,urlcolor=rossos,citecolor=verde}

\newcommand{\TeV}{{\rm ~TeV}}

\newcommand{\mybar}[1]%
        {\kern  .6pt\overline{\kern - .6pt#1\kern - .6pt}\kern  .6pt}
\begin{document}


\title{$Z^{\prime}$ models for the LHCb and $g-2$ muon anomalies}


\author{Ben Allanach}
\email{B.C.Allanach@damtp.cam.ac.uk}
\affiliation{DAMTP, CMS, Wilberforce Road, University of Cambridge, Cambridge, CB3 0WA, United Kingdom}
\author{Farinaldo S. Queiroz}
\email{queiroz@mpi-hd.mgd.de}
\affiliation{Max-Planck-Institut fur Kernphysik, Postfach 103980, 69029 Heidelberg, Germany}
\author{Alessandro Strumia}
\email{alessandro.strumia@cern.ch}
\affiliation{Dipartimento di Fisica dell' Universita di Pisa and INFN, Italy\\
National Institute of Chemical Physics and Biophysics, Tallinn, Estonia}
\author{Sichun Sun}
\email{sichun@uw.edu}
\affiliation{Institute for Advanced Study, Hong Kong University of Science and Technology, Clear Water Bay, Hong Kong}

\date{\today}

\begin{abstract}
We revisit a class of $Z'$ explanations of 
the anomalies found by the LHCb collaboration in $B$ decays, and
 show that the scenario is tightly constrained by a combination of constraints: 
 (i) LHC searches for di-muon resonances, 
(ii) pertubativity of the $Z'$ couplings;
(iii) the $B_s$ mass difference,  and 
(iv) electro-weak precision data.
Solutions are found by suppressing the $Z'$ coupling to electrons and to light quarks and/or by allowing
for a $Z'$ decay width into dark matter.
We also present a simplified framework where a TeV-scale $Z^{\prime}$ gauge
boson that couples to standard leptons as well as to new heavy vector-like leptons, 
can simultaneously accommodate the LHCb anomalies and the muon $g-2$ anomaly.
\end{abstract}

\maketitle

\section{Introduction}

The Standard Model (SM) has been tested at high precision and proven to be the best
description of electroweak and strong interactions. However, we have several
observational reasons to believe that the Standard Model is incomplete,
for example the 
inference of non-zero
neutrino mass and dark matter, the measurement of the muon anomalous
magnetic moment and the results of some recent searches for
new physics, as well as some more fundamental matters such as the
hierarchy problem. 

Many SM extensions have been proposed and often share the
presence of an extra ${\rm U}(1)' $ gauge symmetry. For instance, superstring theory
and grand unification theories provide several examples
\cite{Cvetic:1997ky,Lopez:1996ta,Donini:1997yu}. In supersymmetric grand
unification theories, the ${\rm U}(1)'$ and SM electroweak breaking scales are
usually tied to the soft supersymmetry breaking scale
\cite{Cvetic:1997ky,Keith:1997zb,Mohapatra:1996in}. The TeV focused composite
Higgs and Little Higgs models naturally have a ${\rm U}(1)'$ extension. The recently
constructed \textquotedblleft Little Flavor\textquotedblright~theory
\cite{Sun:2013cza,Sun:2014jha,2015arXiv150905758G} extends both gauge and
fermion generations, connecting them with ``Little Higgs", so as to provide more
experimentally allowed flavor off-diagonal options in both the SM and its
extensions. 

With the advent of the Large Hadron Collider (LHC) those models have been
scrutinized, excluding large regions of parameter space
\cite{Ricciardi:2015iwa}. On the other hand, relatively old 
discrepancies between data and SM predictions (such as
the anomalous magnetic moment of the muon ($g-2$) or 
anomalies observed in
the flavor sector
\cite{Aaij:2013qta,Descotes-Genon:2013wba,Altmannshofer:2013foa,Descotes-Genon:2015xqa,Altmannshofer:2015sma,Beaujean:2013soa,Buras:2013qja,Gauld:2013qba},
new physics searches in di-lepton \cite{Aad:2015owa} and di-boson channels
\cite{Aad:2015owa}) promoted some 
investigations to test whether 
models of new physics are capable of accommodating them. In this work, we will
discuss how one can accommodate the anomalies in a simplified framework,
focusing primarily on resolving the flavor anomalies and $g-2$. The latter
 measurement 
exhibits $3.6\sigma$ evidence for new physics contributions to the muon
anomalous 
magnetic moment which has been measured with high precision
\cite{Bennett:2002jb,Bennett:2004pv,Brown:2001mga,Bennett:2006fi}. 
However, there are sizable uncertainties surrounding the hadronic corrections
to $g-2$ 
\cite{Carey:2009zzb}. 
Thus, it is naive to take the reported
deviation at face value in the light of such large theoretical errors. 
One could imagine taking
different conservative approaches. For instance, one could try to accommodate
the measured 
value at the $2\sigma$ level, or one could derive limits on the
mediator/couplings involved in the $g-2$ loop diagram by assuring that the
contribution is smaller than the uncertainty in the measurement. Here we will
take the former 
approach and comment on possible bounds if the latter approach had been
followed.  

Moreover, flavor physics poses some intriguing questions. Rare $B$ decays
mediated by the flavor-changing neutral $b\rightarrow s$ transition are
sensitive probes for beyond the Standard Model (BSM) physics. The decay $B
\rightarrow K^*(\rightarrow 
K\pi) \mu^+\mu^-$ has a few observable quantities including the branching
ratio and the angular distribution of its four-body final state. 
The measurement has been performed at $B$ factories~\cite{Wei:2009zv}, 
Tevatron experiments~\cite{Aaltonen:2011qs,Aaltonen:2011ja},
LHCb~\cite{Aaij:2013iag}, 
ATLAS~\cite{Aaij:2013iag} and CMS~\cite{CMS:cwa}. 
According to some authors,
the LHCb measurements contain deviations from SM expectations which require
non-trivial explanation, since several observables in $B \rightarrow
K^*(\rightarrow K\pi) \mu^+\mu^-$ as well as in other decays such as
$B_s\rightarrow 
\mu^+\mu^-,B\rightarrow K \mu^+\mu^-,B\rightarrow X_s \gamma $ agree with SM
predictions, within uncertainties. Moreover the low value of recent measurements  $R_K=\textrm{BR}(B \rightarrow K \mu^+ \mu^-)/\textrm{BR}(B \rightarrow K e^+ e^-)$ suggest beyond the standard model lepton non-universality.

Hence, we have exciting and puzzling signals at our disposal but the situation
 is far from clear, and attempts that account for one anomaly at time in the
 context of a heavy $Z'$ boson have been put forth
 \cite{Descotes-Genon:2013wba,Altmannshofer:2013foa,Buras:2013qja,Crivellin:2015xaa,extraZ'}. Here we
 revisit such $Z'$ scenarios using a search for di-muon resonances
 from ATLAS and perturbativity of the $Z'$ coupling, 
 along with updated constraints on the $B_s$ mass 
 difference to show that a class of $Z^{\prime}$ models motivated by the LHCb anomaly are  disfavored by data,
 unless the $Z'$ is rather heavy and strongly coupled.
 Solutions are found by suppressing the $Z'$ coupling to electrons and to light quarks and/or by allowing
for a $Z'$ decay width into dark matter.
Values of the $Z'$ mass, $M_{Z'}\approx 1.9\TeV$, suggested by the diboson anomaly become allowed.

Lastly, we propose a simplified model capable of 
simultaneously addressing 
the discrepant anomalous magnetic of the muon and 
flavor
physics anomalies in a similar vein to Refs.~\cite{Ricciardi:2015iwa,Sierra:2015fma,Belanger:2015nma,Gripaios:2015gra}. 
The model evades other existing limits from precision
flavor physics while predicting interesting LHC phenomenology.



\section{$B \rightarrow K^*\mu^+\mu^-$ Anomaly in $Z^{\prime}$ Models \label{sec:lhcbanom}}

Recent LHCb measurements of the angular distributions in the $B \to
K^\ast \mu^+ \mu^-$ decay
and the low value of
$R_K=\textrm{BR}(B \rightarrow K \mu^+ \mu^-)/\textrm{BR}(B \rightarrow K e^+ e^-)$ 
suggest  deviations from Standard Model
(SM) expectations \cite{Aaij:2013qta}. 
New physics can fit such anomalies, provided that it generates the following effective operator
\begin{equation} 
\label{Eq1}
 {\cal L} \supset  \frac{4 G_F}{\sqrt{2}} \, \frac{\alpha}{4 \pi} \,
 V_{ts}^\ast V_{tb} \, \Delta C_9  \left (\bar s \gamma_\alpha P_L b \right )
 (\bar \mu \gamma^\alpha \mu) + {\rm h.c.}, 
\end{equation} 
with coefficient 
$\Delta C_9 \approx -1.07\pm 0.26$~\cite{Altmannshofer:2013foa,Hiller:2014yaa,Descotes-Genon:2013wba,newer}.

\smallskip

\begin{figure}[!t]
\includegraphics[scale=0.25]{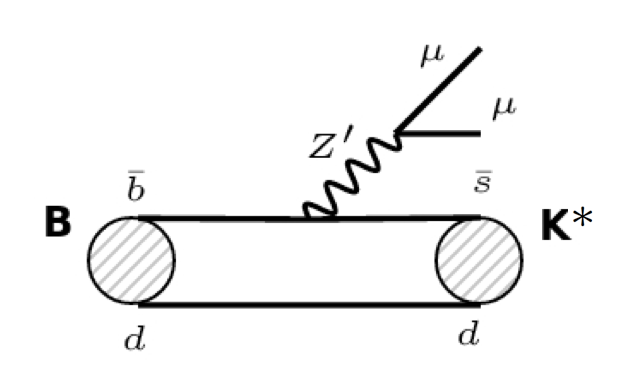}
\caption{$Z'$ contribution to $B \to
K^\ast \mu^+ \mu^-$ decay} 
\label{fig:BS}
\end{figure}

In order to generate this operator, Ref.~\cite{Gauld:2013qba} considered a `toy model'
where the Standard Model is extended by adding a massive $Z'$ that couples to leptons and to $\bar b s$.
Going to more firm theoretical grounds, we extend the SM gauge group adding one extra abelian 
factor ${\rm U}(1)_X$, which
introduces a massive $Z^{\prime}$ boson.
If the ${\rm U}(1)_X$ charges of the left-handed quark doublets are flavor-dependent,
the $Z'$ acquires flavor-violating coupling to left-handed down quarks, 
assuming that the CKM matrix mostly comes from rotating down quarks to their mass-eigenstate basis
(this is a plausible assumption given that mass hierarchies are large in the up-quark sector than in the down sector,
typically leading to smaller mixings).
Motivated by these considerations and by Eq.~(\ref{Eq1})
we thereby consider the following minimal Lagrangian:
\begin{eqnarray} 
{\cal L} & \supset  & \frac{Z^{\prime \mu}}{2\cos\theta_{\rm W}} \bigg[ \gV 
   ( \bar \mu \gamma_{\mu} \mu +  \bar \nu_\mu \gamma_{\mu} P_L  \nu_\mu ) +  \nonumber \\
&&  g_t (\bar t \gamma^{\mu} P_L t+\bar b\gamma^{\mu} P_L b)
 +\gbs  \sum_{q } (\bar{q}\gamma^{\mu} P_L q)+  \label{Eq3} \\
 && +(g_t-\gbs)  ( V_{ts}^\ast V_{tb}\, \, \bar s \gamma_{\mu} P_L\, b + {\rm h.c.})\bigg]\nonumber
\end{eqnarray}
where $P_L = (1-\gamma_5)/2$\, projects over left-handed fields and the sum runs over $q=\{u,d,s,c\}$.
We assumed a common 
$Z'$ coupling $\gbs $ to  1st and 2nd generation left-handed quarks
(in order to avoid large flavour violation among light quarks), 
a coupling $g_t$ to 3rd generation left-handed  quarks $t,b$, 
and one vectorial coupling $\gV $ to muons.
Then, the coupling to $\nu_\mu$ arises because of SU$(2)_L$ invariance, 
and the coupling to $\bar b s$ (as well as similar terms, that have been omitted)
arises after performing the CKM rotation to mass eigenstates in the down sector.
Fig.~\ref{fig:BS} shows a process whereby these effective $Z'$ couplings can
fit the $B$ anomalies 
generating the effective operator in Eq.~(\ref{Eq1}) with coefficient
\begin{equation} \label{EqC9} 
\Delta C_9 = -\frac{\pi g_\mu (g_t-g_q)}{2\sqrt{2}G_{\rm F} M_{Z'}^2\alpha
  \cos^2\theta_{\rm W}} 
 \,. 
\end{equation}

\subsection{Constraints}

Here we discuss general limits on neutral vector bosons that are of interest to the LHCb anomaly.

\subsubsection{$B_s$ mass difference}

The existence of a massive $Z^{\prime}$ gauge boson alters the prediction of
the mass difference ($\Delta M_{B_s}$) of the $B_s$ meson, whose deviation
from the SM expectation can be approximated as~\cite{Gauld:2013qba}
\begin{equation} \label{Eq5}
\Delta_{B_s} \simeq 3.1 (g_t-\gbs) ^2 \left(\frac{{\rm TeV} }{M_{Z^\prime}}\right)^2\left [ 1 - 0.029 \, \ln  \frac{M_Z^\prime}{{\rm TeV}}  \right ] ,
\end{equation}
where $\Delta_{B_s} \equiv  \Delta M _{B_s}^{Z^\prime}/\Delta M_{B_s}-1$,
$M_{Z^\prime}$ is the $Z^{\prime}$ mass and $g_t-g_q $ is the coupling 
between $Z'$ and $b, s$ quarks implied by Eq.~(\ref{Eq3}).

Current measurements impose the upper limit 
\begin{equation} \label{Eq10}
|\Delta_{B_s} | < 8.4\% 
\end{equation} 
 at 95\% confidence level~\cite{GlobalfitBs}.
The resulting bound is shown in Fig.~\ref{figdimuon}, and
is slightly stronger than in Ref.~\cite{Gauld:2013qba}, where the outdated
bound $|\Delta_{B_s} | < 20\% $ was used.

Having assumed that first generation quarks have a common $Z'$ charge,
the ratio between the $Z'$ correction to $\epsilon_K$ and to $\Delta M_{B_s}$ 
is comparable to the SM prediction for the same ratio.
$\epsilon_K$ then does not lead to an obviously more stringent constraint
than the $\Delta M_{B_s}$ constraint.
Given that $\epsilon_K$ involves extra model-dependent issues, we ignore it in
the following.

\subsubsection{Electro-weak precision data}
The $Z'$ gives rise to various
corrections to electro-weak precision observables.
Had we considered a $Z'$ coupled to electrons (with a coupling $g_e$ comparable to $g_\mu$, like
in Ref.~\cite{Gauld:2013qba})
observables measured at LEP with per mille precision would have been affected
at tree level, giving rise to bounds of the form $g_e^2 M_Z^2/M_{Z'}^2 <
10^{-3}$, too strong for our purposes.
We instead assumed a $Z'$ that does not couple to electrons nor to the Higgs
doublet, 
such that it is very weakly constrained and its phenomenology is similar to
the muonphilic $Z'$ studied in~\cite{Salvioni}. 

\smallskip

The only observable affected at tree level is $\nu_\mu/$nucleon scattering, measured with per cent accuracy by the
NuTeV collaboration, which claimed an anomaly in the (neutral current)/(charged current)
ratio of deep inelastic $\nu_{\mu}$-nucleon scattering~\cite{Zeller:2001hh}. 
If this anomaly is not due to underestimated SM uncertainties,
new physics can fit the NuTeV anomaly, provided that it generates the following effective operator~\cite{Davidson:2001ji}
\begin{equation} 
\label{EqNuTeV}
 {\cal L} \supset (-38\pm14) \frac{4 G_F}{\sqrt{2}} \frac{\alpha}{4\pi} \, 
(\bar Q \gamma^\alpha Q) 
(\bar \nu_\mu \gamma_\alpha P_L \nu_{\mu}  )
\end{equation} 
where $Q$ are the SU(2)$_L$ left-handed quark doublets.
Altough this operator  has a structure analogous to the one suggested by the LHCb anomaly, see eq.~(\ref{Eq1}),
the coefficient in Eq.~(\ref{EqNuTeV}) is significantly larger.
Thereby  the $Z'$ motivated by the LHCb anomaly generates this operator with a
coefficient smaller than what is needed to fit the NuTeV anomaly,
at least unless one assumes $|g_t - g_q|\ll g_q$.
Viewing NuTeV data as a one-sided bound, it is safely satisfied in the parameter
region of our model which successfully explains the LHCb anomaly.


\medskip

The precision observables measured with greater than per-mille accuracy are
affected only at the loop level. 
Among them, the most precise measurements are those in the lepton sector. 
The $Z'$  affects the $Z\to \mu\mu$ width as well as the relative forward-backward asymmetry, measured at LEP.
Given that the bound is relatively weak, it is enough for our purposes to estimate it as
\begin{equation} 
\label{Eq7bis}
\Delta_{\mu}  \sim \frac{g_\mu^2}{(4\pi)^2} \frac{M_Z^2}{M_{Z'}^2} < 10^{-3}.
\end{equation}
Furthermore, the anomalous decay moment of the muon (discussed later), and the $\mu$ decay rate
(considered in Ref.~\cite{Gauld:2013qba} as a `CKM unitarity bound')
only receive corrections proportional to $m_\mu^2/M_{Z'}^2$.


%
%
%

\subsubsection{ATLAS di-muon resonance search}

\begin{figure}[!t]
\includegraphics[scale=0.6]{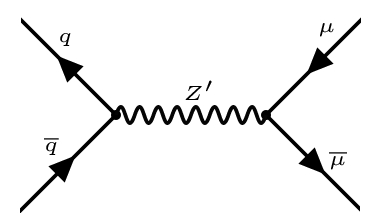}
\caption{$Z'$ contribution to LHC di-muon resonance
  production. \label{fig:dimuon}} 
\end{figure}

\begin{figure}[!t]
\includegraphics[width=\columnwidth]{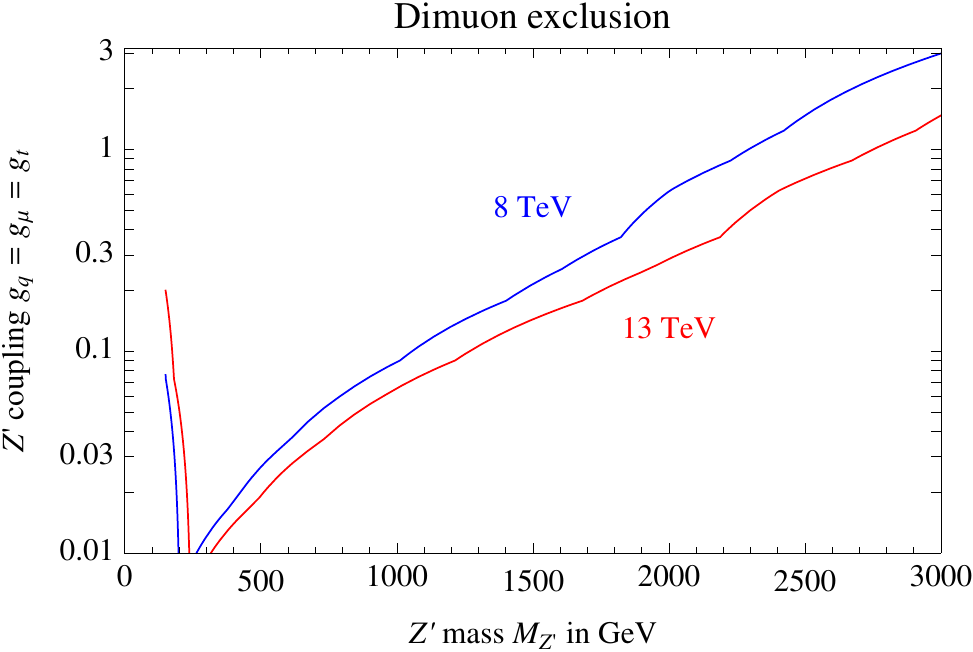}
\caption{Upper bound on the $Z'$ couplings as function of the $Z'$ mass.
We  used  di-muon resonance search data from an 8 TeV $20^{-1}$ fb ATLAS
analysis~\cite{Aad:2014cka} and from 13 TeV data. } 
\label{figdimuon}
\end{figure}

Searches for di-lepton resonances at the LHC have proven to be an excellent
probe of models that predict new neutral vector bosons that have sizeable
coupling to leptons
\cite{Alves:2015pea,Patra:2015bga,Patra:2015bga,Alves:2015mua,Arcadi:2014lta, 
Chen:2015tia,Marcos:2015dza,Kahlhoefer:2015bea,Baker:2015qna}. In particular, bounds on the
mediator mass coming from di-lepton searches are more stringent than those
from dijets searches in most models due to a reduced background, except for
models in which the new neutral vector bosons are leptophobic
\cite{Alves:2013tqa,Duerr:2014wra,Perez:2015rza,An:2012ue,Perez:2014qfa}. In order to
evaluate the constraints from the 20 fb$^{-1}$, 8 TeV ATLAS data
\cite{Aad:2014cka} and from 13 TeV first run data \cite{LHC13} with 3.2fb$^{-1}$. The dilepton invariant mass
spectrum is the discriminating factor in both searches, and since no significant deviations from the
Standard Model expectation has been observed restrictive bounds were placed.
we implemented the model in {\tt
  Feynrules2.0}~\cite{Alloul:2013bka} and later used {\tt Madgraph5}~\cite{Alwall:2014hca} to simulate 
signal events. We simulated di-muon pair production with up to one extra 
jet, 
and accounted for showering, hadronization, detection effects and jet
clustering using {\tt Pythia8.212}~\cite{Sjostrand:2014zea} and {\tt
  Delphes3.0}~\cite{deFavereau:2013fsa} packages and compared with the
background 
reported in Ref.~\cite{Aad:2014cka,LHC13}.

  Fig.~\ref{figdimuon} displays the upper bound on $g_\mu$ as a function of
  $M_{Z'}$ 
  for $g_\mu = g_q=g_t$.  
  The bound can be rescaled to generic values of the couplings $g_\mu,g_q,g_t$ taking into account that,
  in the narrow width approximation, the signal rate scales as 
  \begin{equation} 
\sigma(pp\to Z'\to \mu\mu)\propto \frac{g_q^2 g_\mu^2}{ g_\mu^2+4 g_q^2+2 g _t^2 }.
\end{equation}

\begin{figure*}[t]
\centering
\mbox{\includegraphics[width=0.66\columnwidth]{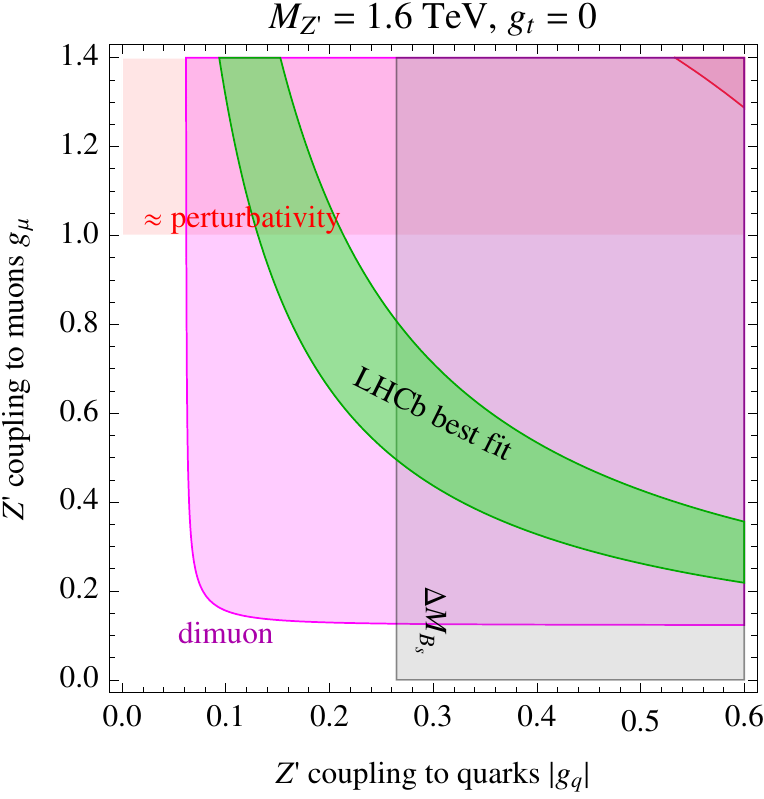}~~
\includegraphics[width=0.66\columnwidth]{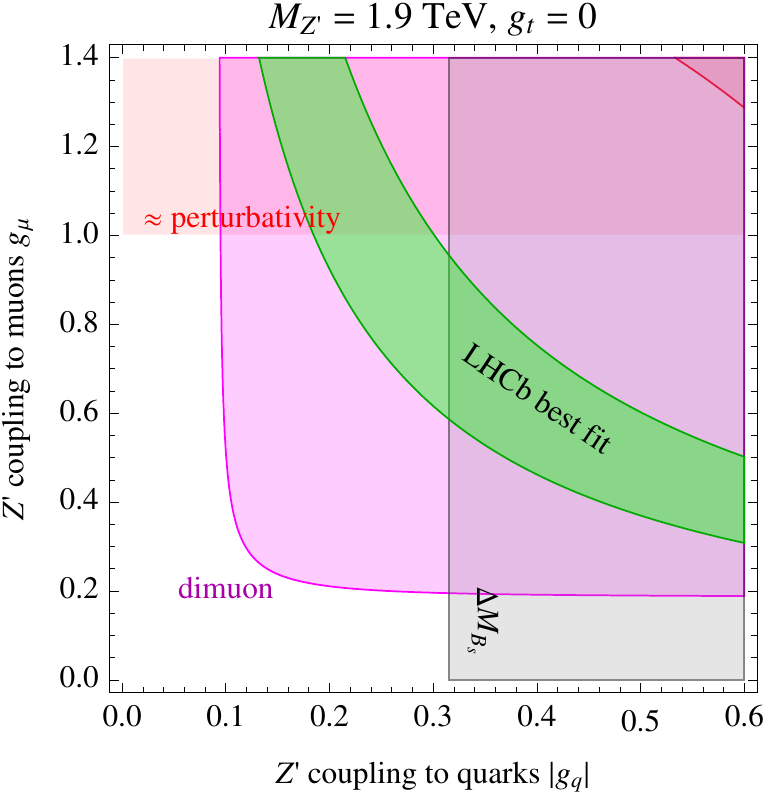}~~
\includegraphics[width=0.66\columnwidth]{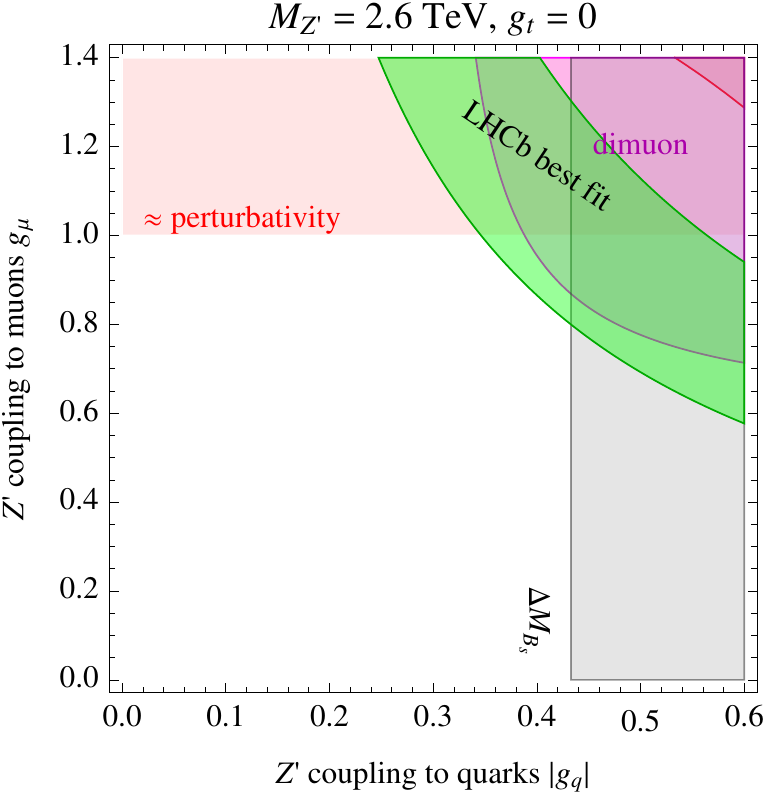}}
\caption{Region in the plane ($g_q, g_\mu$) of $Z'$ couplings to quarks and to muons 
that accommodates the LHCb anomaly (in green),  superimposed with limits from
  di-muon data (in magenta), from the $B_s$ mass difference (in gray) and perturbativity (in red)
  for different values of the $Z'$ mass.
  Electro-weak precision data give a weaker bound on $g_\mu$ which does not appear in the panels.}
\label{Graph01}
\end{figure*}

\subsubsection{Perturbativity}
Imposing that the gauge coupling of a generic $Z'$ can be extrapolated up to the Planck energy
without hitting Landau poles implies an upper bound on its width
\begin{equation}\label{GammaM}
\frac{\Gamma_{Z'}}{M_{Z'}}  < \frac{\pi}{2} \frac{1}{\ln M_{\rm Pl}/M_{Z'}} \approx 0.04.
\end{equation}
This model-independent condition holds because the same diagram generates both
the real part of the $Z'$ propagator (that dictates the renormalization group running)
and the imaginary part of the $Z'$ propagator (that describes the width), with a universal relation among them.
New physics at very large energy can affect this bound, making it stronger if extra scalars or fermions charged under the U(1) are present,
and weaker if extra vectors are present such that the U(1) gets embedded in a non-abelian group.

In the case at hand we have
\begin{equation}
\frac{\Gamma_{Z'}}{M_{Z'}} = \frac{g_\mu^2+4 g_q^2+2 g _t^2}{8\pi(2\cos\theta_{\rm W})^2}.
\end{equation}
This only excludes the upper corners (in red) of the panels in Fig.~\ref{Graph01}.

Furthermore, as an extra naive and semi-quantitative but reasonable perturbativity condition, 
we impose that the $Z'$ gauge couplings to muons,
$\gV$,  and to quarks, $\gbs $ (precisely defined by Eq.~(\ref{Eq3})),
 must be smaller than $\approx 1$. 
This excludes the region in orange in Figs.~\ref{Graph01}-\ref{Graph02}.
The reader should keep in mind that this requirement is reasonable but the
precise value of the perturbative bound is arbitrary. 
New physics effects that accommodate the LHCb anomaly can be obtained within the perturbative regime~\cite{Altmannshofer:2013foa}. 
To the best of our knowledge there is no strongly coupled ${\rm U}(1)$ gauge theory
that arises naturally, see for instance, Left-Right models
\cite{Mohapatra:1974hk,Senjanovic:1975rk,Senjanovic:1978ev,Parida:2012sq,Rodejohann:2015hka}, Little Higgs models
\cite{Sun:2013cza,deAlmeida:2007zzb,deAlmeida:2010qb,Dias:2007pv}, gauged baryon and lepton number models
\cite{Rodejohann:2015lca,Perez:2013tea,Duerr:2014wra,An:2012ue} or 3-3-1 models
\cite{Foot:1992rh,Mizukoshi:2010ky,Cogollo:2012ek,Alvares:2012qv,Alves:2012yp,Caetano:2013nya,
Profumo:2013sca,Kelso:2013nwa,Kelso:2013zfa,Cogollo:2013mga,Cogollo:2014jia,Dong:2014wsa,
Kelso:2014qka}. 


\subsection{Combined analysis}
We are now ready for a combined analysis.


In Figure~\ref{Graph01} we show 
the region favored at $\pm1\sigma$ by the LHCb anomaly (in green),
compared to the exclusion bounds discussed before:
from $\Delta M_{B_s}$ (gray regions), from perturbativity (red regions),
from di-muon data (magenta regions) and from electroweak precision data (the bounds are so weak that they not appear in the plot).
We made plots in the  $(g_\mu, |g_q|)$ plane,   assuming $g_t=0$  and for a
few representative values of $M_{Z'}$. 
In the middle panel we considered $M_{Z'}=1.9\TeV$, a value for which a small $Z'\to ZZ$ or $Z'\to W^+W^-$
decay width (as in \cite{Sun:2014jha}) might fit the diboson
anomaly~\cite{Aad:2015owa}.

Compared to  earlier analyses~\cite{Gauld:2013qba} we used new and updated data,
we included di-muon data, we used a different and more solid theoretical framework,
we assumed that the $Z'$ does not couple to electrons (which allows us to fit
the $R_K$ anomaly) and in order to avoid strong electroweak bounds. 


These new constraints, and in particular the inclusion of LHC di-muon data,
implies that the $Z^{\prime}$ models cannot fit the LHCb anomalies, unless the $Z'$ is heavier than
about 2.5 TeV.  
In such a case, the $Z'$ should be seen in early run II data.
Heavier $Z'$ bosons need larger gauge couplings, which raises issues with
perturbativity. 




\medskip

In view of the strong impact of di-muon data, let us discuss how they can be weakened.
The ATLAS di-muon data put limits on the quantity
$\sigma(pp\to Z') \times {\rm BR}(Z^{\prime} \rightarrow \mu\mu)$.

\begin{figure*}
\centering
\mbox{\includegraphics[width=0.67\columnwidth]{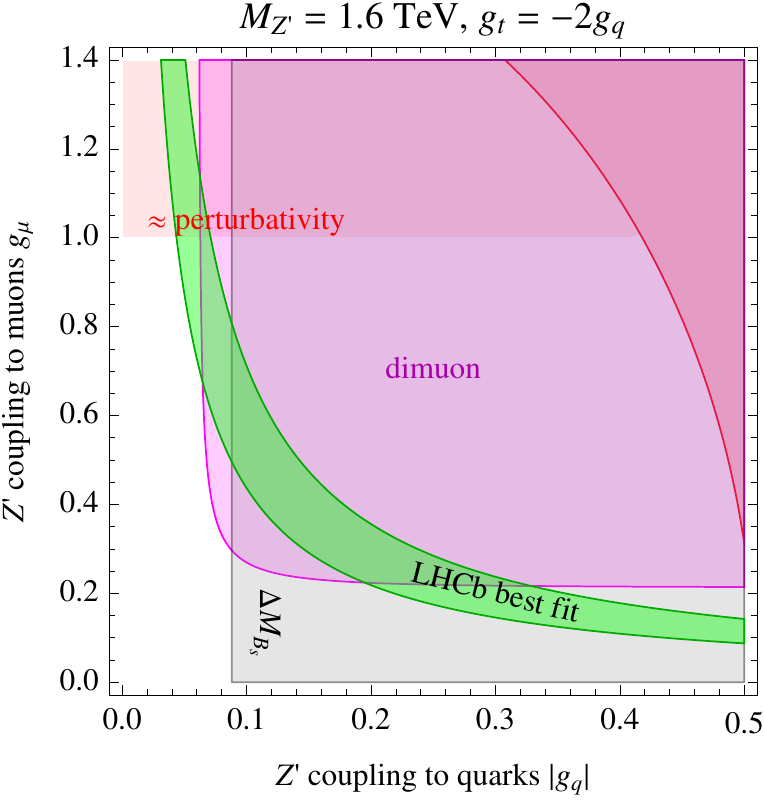}~~
\includegraphics[width=0.67\columnwidth]{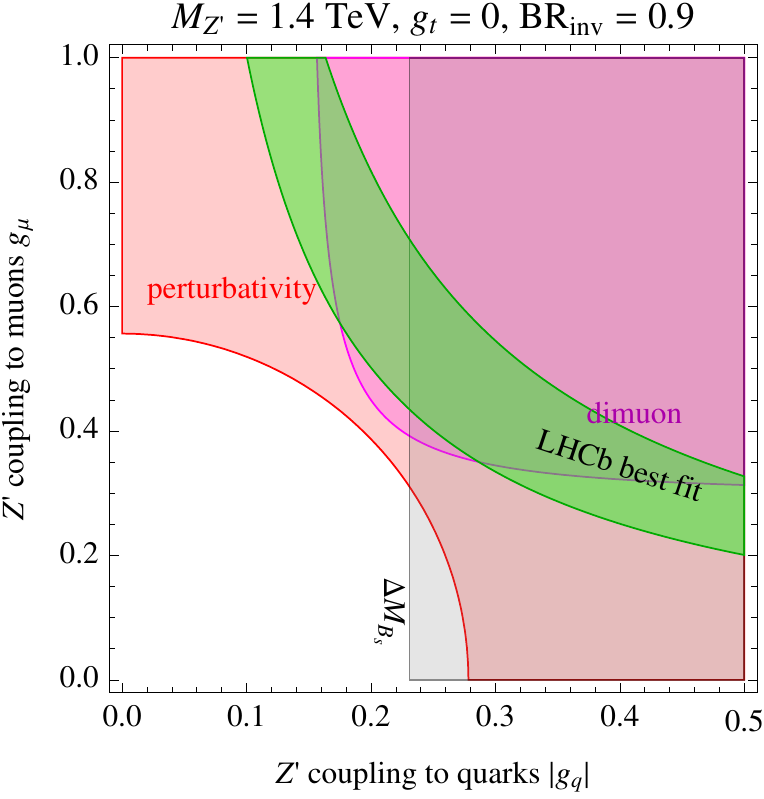}~~
\includegraphics[width=0.67\columnwidth]{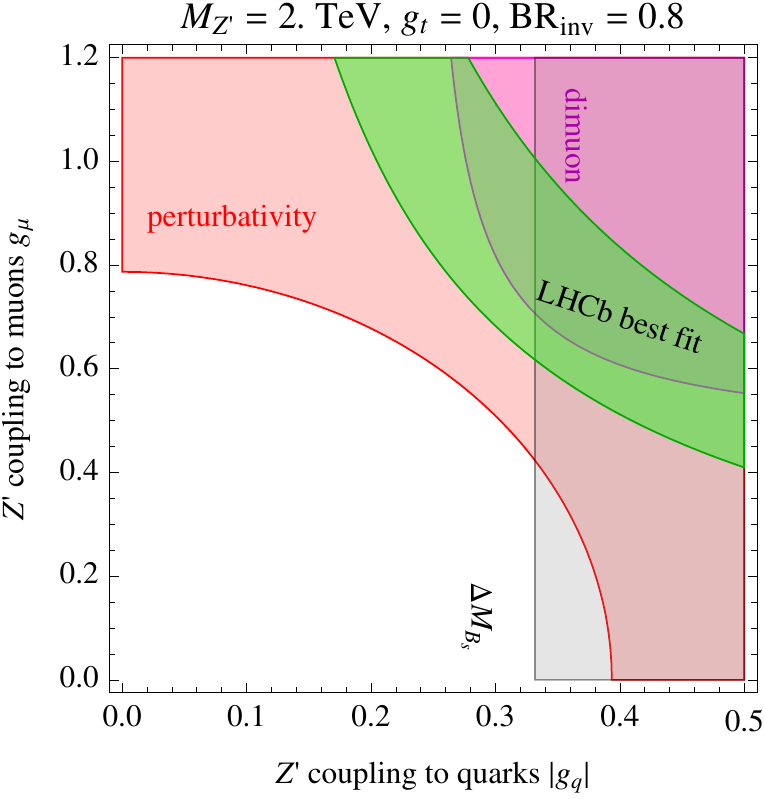}}
\caption{As in Fig.~\ref{Graph01}, but adding extra $Z'$ decay modes into $t\bar t, b \bar b$ (left panel) or into invisible modes, such as dark matter
(middle and right panels).
The presence of extra decay modes suppress the di-muon limits, allowing for global solutions of the LHCb anomalies
for lighter values of the $Z'$ mass.}
\label{Graph02}
\end{figure*}

One way of weakening the di-muon limits consists in reducing $\sigma(pp\to
Z')$, which is proportional to $g_q^2$. 
So far we assumed $|g_t | \ll |g_q|$ i.e.\ a $Z'$ that couples dominantly to light quarks.
In the opposite limit $|g_q|\ll |g_t|$  of a $Z'$ that couples dominantly to third generation quarks,
the $Z'$ coupling to light quarks is CKM suppressed, and the di-muon bound
is no longer constraining.
Fig.~\ref{Graph01} can be reinterpreted as having $|g_t|$ (rather than $|g_q|$) on the horizontal axis by just omitting the di-muon bound.
Furthermore, in the left panel of Fig.~\ref{Graph02} we consider an intermediate situation, $g_t = -2 g_q$, finding that a
$Z'$ with 1.6 TeV mass becomes allowed.

Another way of reducing the di-muon limits consists in keeping $g_q$ sizeable
and reducing ${\rm BR}(Z^{\prime} \rightarrow \mu\mu)$, 
by assuming that the $Z'$ has a sizeable branching ratio into Dark Matter particles, as actually predicted
in some models \cite{Alves:2015mua}.
This situation is explored in the middle and right panels of Fig.~\ref{Graph02}: we see that
global solutions are now allowed for $M_{Z'}$ as light as 1.4 TeV (middle panel).
However, in view of the increased total $Z'$ width, the perturbativity bounds on $\Gamma_{Z'}/M_{Z'}$ of eq.~(\ref{GammaM}) becomes stronger
and start becoming a limiting factor (red regions in Fig.~\ref{Graph02}).



\section{A New Simplified Model for LHCb and Muon $g-2$ Anomalies \label{sec:simp}}

\subsection{Framework and Fit to the Anomalies}
The muon magnetic moment has been measured in the Brookhaven
laboratory, which used a ring of polarized muon beams
\cite{Brown:2001mga,Bennett:2006fi}. The experiment was able to reach a
precision with unprecedented sensitivity which intriguingly resulted into a
$3.6\sigma$ excess over the standard model prediction with $\Delta
(g-2)/2=(295 \pm 81) \times 10^{-11}$. Theoretical uncertainties in
hadronic corrections (namely hadronic vacuum polarization and
light-by-light scattering) blur the significance of the excess, but the $g-2$
measurement at Fermilab along with a concerted effort to improve 
the accuracy of the SM theoretical
prediction should decisively clarify the anomaly.  

Motivated by this long standing excess and the LHCb
anomaly discussed above, we consider a model that extends 
${\rm SU}(2)_L\otimes {\rm U}(1)_Y$ with 
an extra abelian gauge group ${\rm U}(1)_X$ and two additional lepton fields: a
vector-like lepton electroweak doublet $(\nu',\ell')$ and vector-like singlet
$\mu'$. The left-handed part of $\ell'$, $\ell'_\text{LH}$, and the right-handed part
of $\mu'$, $\mu'_\text{RH}$, are charged with the quantum numbers
$(2,-\half,-x)$ and $(1,-1,-x)$ under ${\rm SU}(2)_L\otimes {\rm U}(1)_Y \otimes {\rm U}(1)_X$,
respectively. We 
redefine a SM-like Dirac fermion $L=(\ell'_\text{LH}, \mu'_\text{RH})^T$. $L$
can be very heavy since it mostly gets its mass from the vector-like particle
mass term 
$M_1\bar{\ell'}\ell'+M_2\bar{\mu'}\mu'$. Note that both the SM fermions and new
leptons are charged under the new ${\rm U}(1)_X$. For simplicity we assume here that
the $Z'$ gets mass from a different scalar rather than the SM Higgs, so the SM
Higgs does not have ${\rm U}(1)_X$ charge. This assumption also forbids the new $Z'$
gauge boson mass eigenstate from mixing with the SM $Z$ mass eigenstate at
tree level.  Notice that it still allows the new lepton doublet and lepton
singlet to
have potential Yukawa couplings to the SM Higgs doublet. This leads to
deviations in 
electroweak observables, but these can be easily avoided by a
suppression of the vector-like Yukawa couplings. 

Consequently, the new charged fermion $L$ can have  a purely vector-like coupling to
the $Z^{\prime}$ since the left-handed and right-handed fermions have the same
$Z'$ charges and the Lagrangian piece relevant for the muon magnetic moment is
\beq
\nonumber&& {\cal L}_\text{eff} \supset \bar{\mu} \gamma^\mu  (g_{v} -\gamma_5 g_{a})L  Z_\mu^{\prime}+ h.c.\\
&&+  \bar{\mu} \gamma^\mu  (g^{\prime}_{v  \mu} -\gamma_5 g^{\prime}_{a \mu}) \mu Z_\mu^{\prime}.
\label{eq8}
\eeq
In order to account for the LHCb anomaly, a vector coupling to muons is
needed. Such a setup can be achieved in several extensions of the SM by
simply choosing the left and right-handed field components to transform
similarly under the action of ${\rm U}(1)_X$ (see Table I of \cite{Alves:2015mua}
for explicit 
examples). Under these assumptions, the $Z^{\prime}$ possesses only
vectorial couplings to the exotic charged lepton and the muon, i.e.\ $g_{a
 }\equiv 0$. Notice that there are two diagrams giving rise to corrections to
$g-2$, as shown in Fig.~\ref{Graph02}. One is a diagram containing one new
particle 
only, the $Z^{\prime}$, and 
another with the exotic charged leptons and the $Z^{\prime}$ running in the
loop. We have explicitly checked that the $Z^{\prime}$ correction to $g-2$ is
negligible compared to the one involving the new charged lepton. The result is
\begin{figure}[!t]
\includegraphics[scale=0.5]{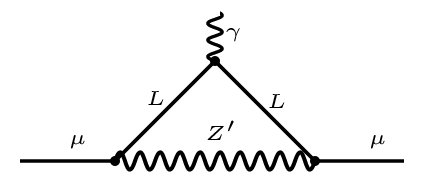}
\includegraphics[scale=0.5]{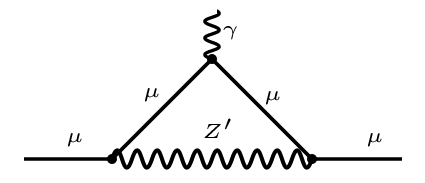}
\caption{Diagrams contributing to the muon anomalous magnetic moment. {\it Upper:}\/
  The dominant contribution to $g-2$ occurs when both the new charged lepton
  and the $Z^{\prime}$ boson run in the loop. 
{\it
    Lower:}\/ The $Z^{\prime}$ correction which is subdominant. \label{fig6}}
\end{figure}
\begin{eqnarray}
&&
\Delta a_{\mu} ({ L}) = \frac{1}{8\pi^2}\frac{m_\mu^2}{ M_{Z^{\prime}}^2 } \int_0^1 dx \frac{g_{v}^2 \ P_{v}(x) + g_{a}^2 \ P_{a} (x) }{(1-x)(1-\lambda^{2} x) +\epsilon^2 \lambda^2 x}\nonumber\\
\label{leptonmuon4}
\end{eqnarray}where
\begin{eqnarray}
P_{v}(x) & = &  2x(1-x)(x-2(1-\epsilon))+\lambda^2(1-\epsilon)^2x^2(1+\epsilon-x) \nonumber\\
\nonumber P_{a}(x) & = & 2x(1-x)(x-2(1+\epsilon))+\lambda^2(1+\epsilon)^2x^2(1-\epsilon-x)\\
\label{leptonmuon5}
\end{eqnarray}where $M_{L}$ is the fermion mass running in the loop, $\epsilon
= M_{L}/m_{\mu}$ and $\lambda= m_{\mu}/M_{Z^{\prime}}$.  In the
$M_{Z^{\prime}} \gg M_{L}$ limit, Eq.~(\ref{leptonmuon4}) becomes
\begin{eqnarray}
&&
\Delta a_{\mu}  ( { L}) = \frac{1}{4\pi^2}\frac{m_\mu^2}{ M_{Z^{\prime}}^2 }\left\lbrace  g_{v}^2\left[\frac{ M_{L} }{m_{\mu}} -\frac{2}{3}\right] + g_{a}^2 \left[ -\frac{ M_{L}}{m_{\mu}} -\frac{2}{3}\right] \right\rbrace. \nonumber\\
\label{leptonmuon6}
\end{eqnarray}

\begin{figure*}[!t]
\centering
\mbox{\includegraphics[width=0.8\columnwidth]{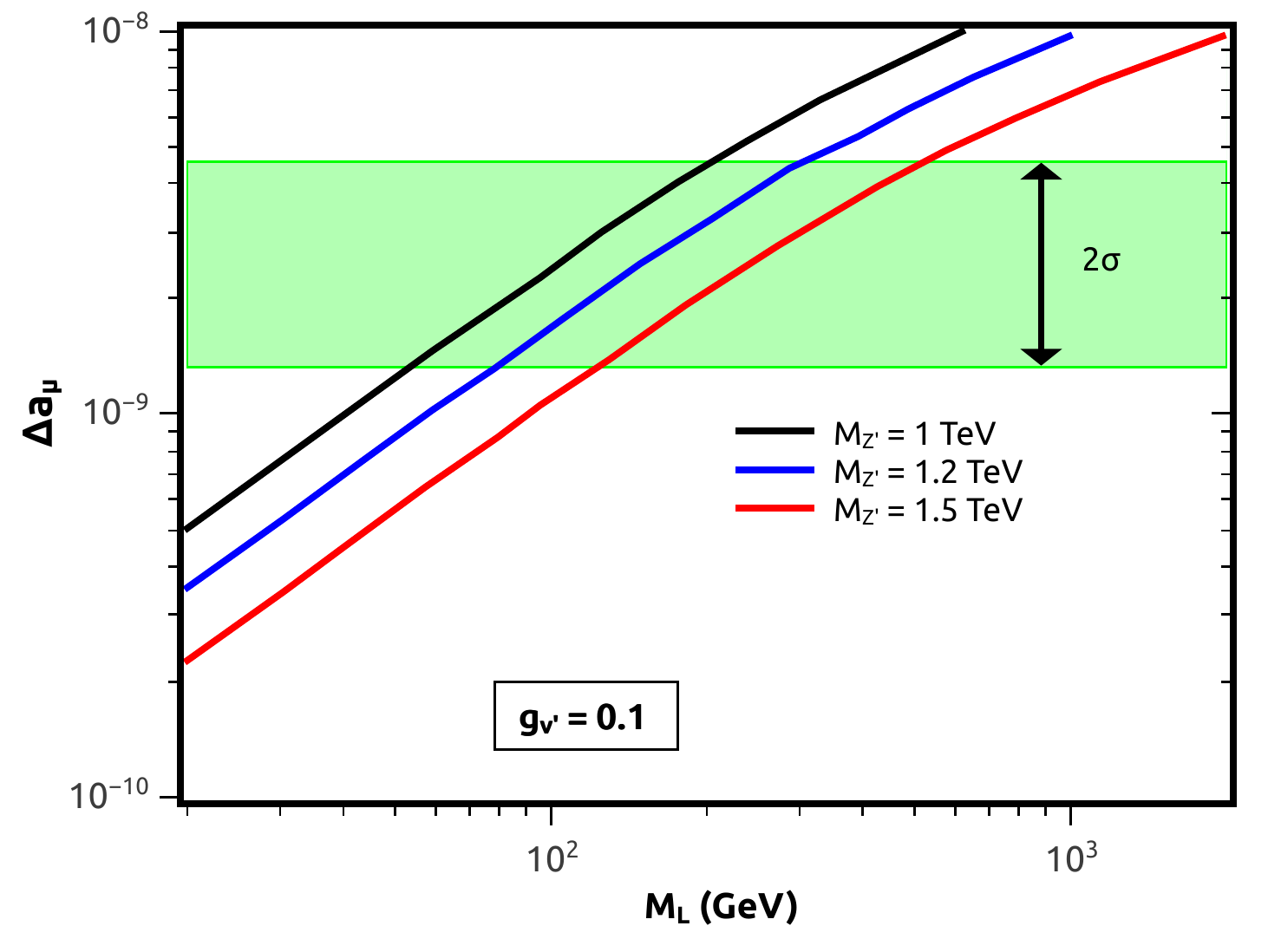}\quad\includegraphics[width=0.8\columnwidth]{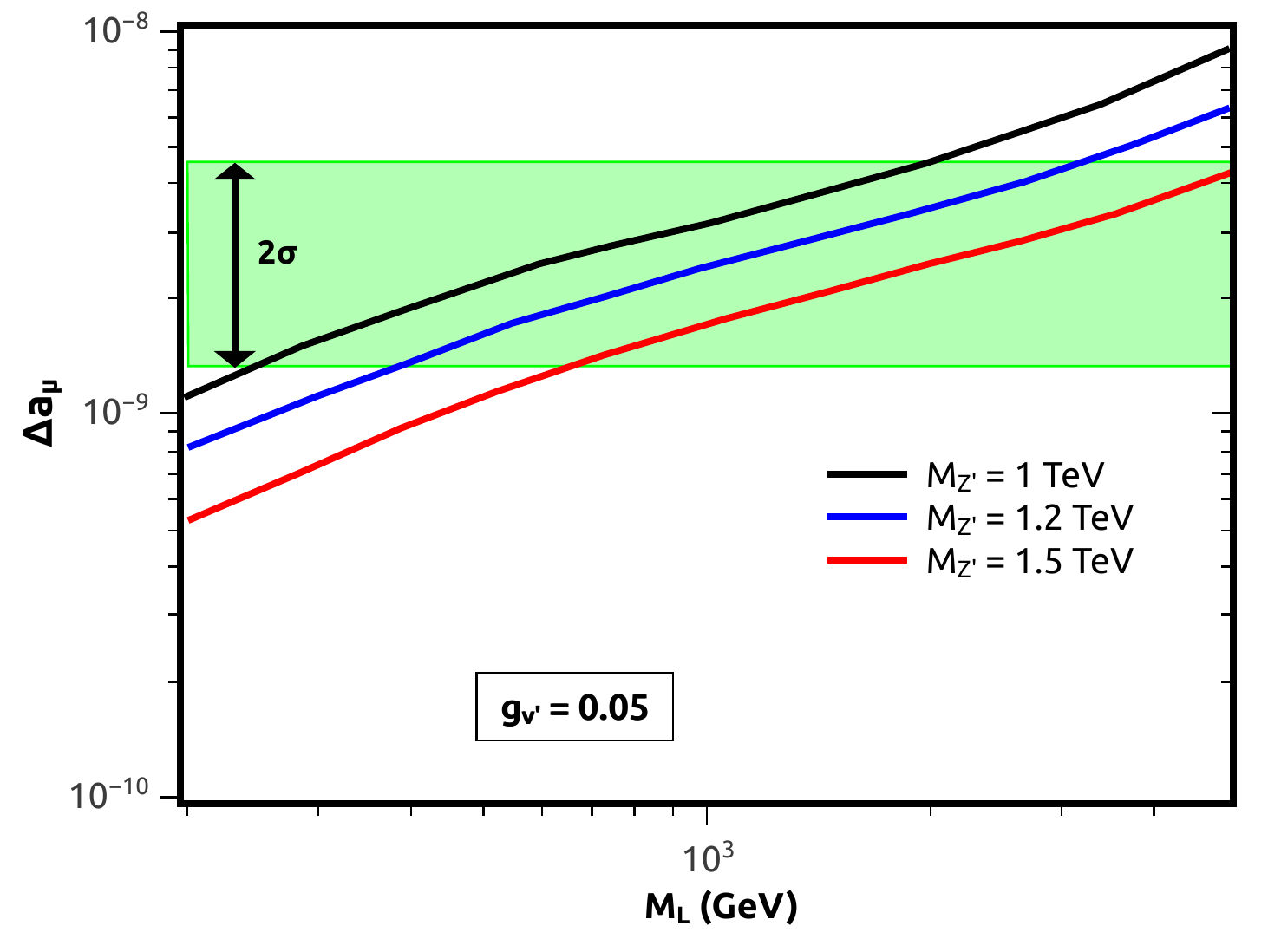}}
\caption{Correction to the muon magnetic moment arising from the presence of
  both the $Z^{\prime}$ and the new charged leptons. It is clear that for
  $g_v=0.1$, one needs a relatively light new charged lepton to
  accommodate $g-2$, whereas for $g_v=0.05$, one can simultaneously
  address the $g-2$ and LHCb anomalies with a TeV scale charged lepton and
  $M_{Z^{\prime}}=1.2\sim 1.8$~TeV, whilst remaining compatible with existing
  limits. A much larger region of parameter space may
  account for the anomalies once other couplings strengths are used.} 
\label{Graph1}
\end{figure*}

It is clear from Eq.~(\ref{leptonmuon6}) that the larger $M_L$ is, the bigger is
the 
correction to $g-2$ from the vectorial coupling. This is due to the fact that
there is a necessary mass 
insertion in the loop correction of $g-2$ to flip the chirality. 
The approximations lose accuracy in the r{e}gime where $M_L$ and $M_Z^{\prime}$ are
comparable and in that case, one should solve Eq.~(\ref{eq8}) numerically
(we did with the help of the Public computer program \cite{Queiroz:2014zfa}). In 
Fig.~\ref{Graph1} we present the contribution to the muon anomalous magnetic
moment 
arising from our model for two specific values of the vector
coupling, $g_v =0.1$ (left panel) and $g_v =0.05$ (right panel), respectively,
for several $Z^{\prime}$ masses. The green band delimits the $2\sigma$ region
that accommodates the muon anomalous magnetic moment. It is clear from
Fig.~\ref{Graph1} that with TeV scale masses we can accommodate the $g-2$
anomaly. 

However, in order to simultaneously address the LHCb and $g-2$ anomalies, dark
matter or visible states should be added in addition to our effective
framework. We have shown that the addition of dark matter or other light
states to a class of $Z^{\prime}$ models opens up a new window to 
accommodate the LHCb anomaly and the inclusion of a exotic charged lepton that
has purely vectorial couplings to the $Z^{\prime}$ gauge boson might foot the
bill. 

Notice that we did not explicitly list couplings
between the $Z^{\prime}$ gauge boson and the dark matter particle, but this
can be easily realised with vector-like Dirac dark fermions, which would play
the 
desired role of relaxing the ATLAS di-muon resonance search constraints.
In the left hand panel of Fig.~\ref{Graph1}, $Z^{\prime}$ decays into dark
matter are not needed, since the $Z^{\prime}$ may decay into new charged leptons
instead. The new charged leptons are relatively light with masses
of 200 GeV or so, within the reach of the next generation of leptonic
colliders. In the right-handed panel of Fig.~7, on the order hand, the
$Z^{\prime}$ is 
not heavy enough to decay into the new charged leptons, whose masses are
mostly at the TeV scale, to explain muon $g-2$ but in agreement with recent
collider limits from the LHC 
\cite{Kumar:2015tna}. In this case the presence of dark
matter or any other light species would be required.  
We will now turn our attention to existing constraints.

\subsection{Experimental constraints}
As we show above, allowing BSM decay modes of the $Z'$ allows it to
be heavier while still explaining the LHCb anomaly. Allowing the $Z' L \mu$
coupling allows the $Z'$ to be heavier while still explaining the g-2 anomaly. 
Because the $Z'$ is allowed to be heavier, other flavor bounds are less
constraining. Thus, the addition of BSM states to the $Z'$ leads to a
framework that is less constrained. 

\subsubsection{$Z \rightarrow 4 \ell$ }

The ATLAS and CMS collaborations  measured the branching ratio of the SM $Z$
decay to four leptons ($Z \rightarrow 4 \ell)$, finding ${\rm BR}(Z \rightarrow 4\ell)=
(4.2 \pm 0.4)\times 10^{-6}$ \cite{TheATLAScollaboration:2013nha}. This
constraint is 
quite restrictive for our class of $Z^{\prime}$ models in the $M_{Z^{\prime}}
< M_Z$ 
mass regime \cite{Altmannshofer:2014cfa}, which is not the case for our
model. 
The new physics contribution to this process that would arise from replacing
the $Z$ in the second vertex by the $Z^{\prime}$ is highly  (as shown) is mass
suppressed. 

\subsubsection{$Z \rightarrow \mu \mu$ }

The precisely measured $Z \rightarrow
\mu\mu$ decay receives corrections from diagrams
with the heavy charged lepton and the $Z^{\prime}$ gauge boson running
in the loop as shown in Fig.~\ref{box2}. This decay is naturally suppressed due
to the TeV scale masses of both charged and $Z^{\prime}$ gauge
boson. 
  
\begin{figure}[t]
\includegraphics[width=0.7\columnwidth]{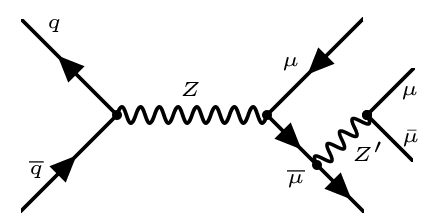}
\caption{The $Z \rightarrow 4\ell$ diagram through a $Z'$.}
\label{box}
\end{figure}

 \begin{figure}[t]
\includegraphics[width=0.5\columnwidth]{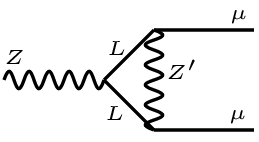}
\caption{The $Z \rightarrow \mu^+\mu^-$ decay diagram via $Z'$.}
\label{box2}
\end{figure}

\subsubsection{Neutrino trident}
\begin{figure}[t]
\includegraphics[width=0.5\columnwidth]{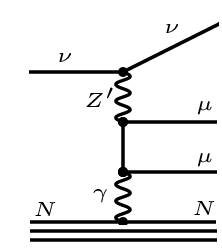}
\caption{Neutrino trident process that leads to constraints on the $Z^{\mu}$ coupling strength to neutrinos-muons, namely $M_{Z^{\prime}}/g_{v\mu} \gtrsim 750$~GeV.}
\label{box3}
\end{figure}

$Z'$ vector fields that only or mostly couple to leptons are constrained by
the process shown in 
Fig.~\ref{box3}~\cite{Altmannshofer:2014pba}. In the heavy mediator regime,
one can write down 
a dimension six effective Lagrangian operator $g^2_{v\mu}( \bar{\mu}
\gamma^{\mu} \mu )(
\bar{\nu}\gamma_\mu P_L\nu)/ M_{Z^{\prime}}^2$ and derive a bound
on a function of the coupling and $Z^{\prime}$ mass, namely
$M_{Z^{\prime}}/g_{v\mu} \gtrsim 
750$~GeV \cite{Altmannshofer:2014pba}. This bound is safely obeyed by our model.

\subsubsection{$\tau$ decay}

\begin{figure}[t]
\includegraphics[width=0.5\columnwidth]{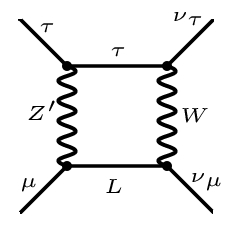}
\caption{One of the box diagramd giving a correction to the $\tau \rightarrow \mu\nu_\tau\bar{\nu_\mu}$ with the new BSM $Z'$,$L$.}
\label{box4}
\end{figure}

The $Z'L\mu$ coupling leads to a correction to $\tau$ decay through the
one-loop box diagram in Fig.~\ref{box4}:
\beq
\frac{{\rm BR}(\tau \rightarrow \mu\nu_\tau\bar{\nu_\mu})}{{\rm BR}(\tau \rightarrow
  \mu\nu_\tau\bar{\nu_\mu})_{\rm SM}} \simeq 1+\delta.
\eeq
Comparing the experimental value~\cite{Beringer:1900zz} 
with the SM prediction for the 
tau lifetime~\cite{Altmannshofer:2014cfa}, we find
\beq
\delta =(7.0\pm3.0)\times 10^{-3}.
\eeq
We can approximate the box diagram contribution from the extra $Z'$,$L$, in
the   $M_L \sim M_{Z'} \gg M_W$ limit as
\beq
\delta \sim \frac{g'_v g}{4\pi^2} \frac{m_W^2}{M_{Z'}^2} .
\eeq
The interaction strength between $W$, $L$ and $\nu$ is fixed by ${\rm SU}(2)_L$
invariance to be $g$, the ${\rm SU}(2)$ gauge coupling strength.
For $\gV ^{\prime} \lesssim 0.1$, our framework satisfies the current
upper limits on $\delta$.  

\subsubsection{$Z'$ decays into new particles}

In our scenario, the $Z'$ decays SM particles, but also into the new lepton $L$, giving
rise to the  decay channels which include
$Z' \rightarrow \mu L$ and $Z' \rightarrow LL$, 
followed by $L \rightarrow W \nu_\mu$. Thus, the $Z^{\prime}$ might produce, in
addition to di-muon resonances, $WW$ plus missing transverse momentum ($p \!
\! \! /_T$), or $W \mu p \! \! \! /_T$ signatures
that we plan to investigate in future. The $W$'s will then subsequently decay
either into either a boosted hadronic $W$-jet or into a lepton plus $p  \! \! \!
/_T$. 

\subsection{Potential ultra-violet completions} 

In our simplified framework, we chose the heavy leptons to be vector-like 
in order to
give them large masses, whilst avoiding constraints on
fourth generation chiral particles. The electroweak precision tests are still
reasonably well satisfied 
since contributions from fermions with opposite chirality
cancel. 

We dub this as an \textquotedblleft  effective $Z'+L$\textquotedblright 
setup, as distinct from a conventional minimal
$Z'$ model~\cite{Langacker:2008yv}. The extra lepton fields are introduced to
potentially incorporate different flavor scenarios. The couplings in this
model (such as family non-universal couplings) can arise in different ways
in explicit models~\cite{Chiang:2006we}, vector-like fermion extensions
\cite{Carone:2013uh}, or considering an effective approach where all $Z'$
couplings are only be generated by higher dimensional operators following the
same arguments as in Ref.~\cite{Fox:2011qd}. We summarize different types of
effective couplings between the $Z^{\prime}$ and fermions below:  
\begin{enumerate}
 \item 
 SM flavor diagonal and off-diagonal $Z^{\prime}$ couplings. They have the potential
 to explain the LHCb anomaly.
 \item
 SM leptons and BSM lepton $Z^{\prime}$ couplings giving rise to the muon anomalous
 magnetic moment deviations. 
 \item
 New heavy lepton couplings to $Z^{\prime}$: these open up new decay channels for
 $Z^{\prime}$
 searches, 
 reducing the decay branching ratio to SM fermions and weakening $Z'$
 leptonic resonance searches. 
\end{enumerate}

The effective approach allows us to treat all three types of couplings as free
parameters, although in most of the UV completed models these parameters will
be related. 
This \textquotedblleft effective $Z'$+L\textquotedblright can arise from 
certain types of \textquotedblleft Little Higgs\textquotedblright model, or
non-minimal supersymmetry. One of the reasons to have extra vector-like
fermions is to decouple the new $Z'$ from the SM fermions. In the
\textquotedblleft Little Flavor\textquotedblright~\cite{Sun:2013cza,Sun:2014jha,2015arXiv150905758G} model, $[{\rm SU}(2)\times
{\rm U}(1)]^2$ breaks down to a diagonal $[{\rm SU}(2)\times {\rm U}(1)]_{\rm SM}$ subgroup via the
little Higgs. The additional ${\rm SU}(2)$ (predicting mixed $ZWW'$ vertices) can
provide 
a possible explanation for the diboson anomaly \cite{Aad:2015owa}. The
SM fermions and extra vector-like fermions are charged under different copies
of the original $[{\rm SU}(2)\times {\rm U}(1)]$ and end up both charged under $[{\rm SU}(2)\times
{\rm U}(1)]_{\rm SM}$. This leaves the SM couplings of fermions and gauge bosons
unaltered, whilst ensuring a skew factor in the BSM
couplings of heavy fermions and gauge bosons. A similar mechanism also
decouples the Kaluza-Klein resonances of gauge bosons and fermions in
Randal-Sundrum 
models. For more discussion on the effective $Z'$ couplings, see
Ref.~\cite{Fox:2011qd}.

\section{Conclusions \label{sec:conc}}

We  revisited the LHCb  anomalies in $B$ decays
in the context of models with an extra $Z^{\prime}$ that couples to
quarks and muons, finding that such an interpretation is tightly constrained by 
the ATLAS di-muon resonance search, 
by the $B_s$  mass difference, and by perturbativity arguments. 
We have shown how the LHCb anomaly can be fitted compatibly with all constraints by a
$Z'$ that dominantly couples to third-generation quarks,
or that has a sizeable branching ratio into dark matter or similar species.

We later proposed an effective $Z^{\prime}$ model with
extra TeV scale vector-like fermions, 
finding that it can reconciles both the measurement of
$g-2$ and the LHCb 
anomaly while obeying existing limits such as those coming from $\tau$-decay,
the neutrino trident process and $Z \rightarrow 4\ell$. 
Such $Z^{\prime}$ is a viable 
option to explain the LHCb and $g-2$ anomalies.


\begin{acknowledgments}
This work was supported by the CRF Grants of the Government of the Hong Kong
SAR under HUKST4/CRF/13G and STFC grant ST/L000385/1. We thank Paolo Ciafaloni,
Hiren Patel, Carlos Yaguna, Branimir Radovcic and Pavel Fileviez Perez  for
illuminating discussions. 

\end{acknowledgments}

\end{document}